A translation of Pierre Marx's 'Bosses du champ électromagnétique:
Modèles électro-gravitationnels de l'électron classique et de l'énergie noire'

# Electromagnetic field bumps
## Electro-gravitational models of the classical electron and dark energy

PIERRE MARX
102, les bois du cerf, 91450 Étiolles, France
marx.p@wanadoo.fr

ABSTRACT. Previously [1] we conjectured that extremely high Electromagnetic (EM) fields in a vacuum generate a gravitational field that causes Maxwell's equations to no longer be linear. This results in a "4-force" in the field configuration space, also called "four-current density". Based on the 4-current density, we postulate that the electric charge is the result of a high EM field. Considering the electrostatic potential with central symmetry and with the assumption that it cannot exceed the Planck potential[1], we show that the associated electric charge is confined inside a sphere with non-zero radius, centered at the origin. We propose to represent this elementary charge *e*, thus replacing the point electron of the standard model with a "bump" of the EM field. According to this model, the gravitational field created by EM is repulsive up to the classical radius of the electron, then attractive beyond. This property provides a model for dark energy, which is believed to cause the accelerated expansion of the universe and whose nature is unknown today. We thus replace the cosmological constant of the standard model ΛCDM by a "field of EM bumps".

**1 Electromagnetic gravitational field**

**1.1 Maxwell's Equations (2nd group)**

The principle of superposition in a vacuum, at the heart of Maxwell's electromagnetic theory and synonymous with the linearity of the eponymous equations, perhaps constitute a way forward, based on the very general fact that an effect cannot be proportional to its cause regardless of its magnitude. Beyond a certain range, nonlinear effects appear, making the usual equations inadequate. This conjecture was the subject of an article published in 2009 in the Annales de la Fondation Louis de Broglie (vol. 34-1) under the title "Gravitational effects of intense electromagnetic fields" [1].

---

[1] In the system of natural units, as described by Planck, the Planck potential has the expression $V_p = c^2 / 4\pi\varepsilon_0 G$

Specifically, it has been assumed that, in a vacuum, the superposition principle ceases to be valid for extremely intense Electromagnetic (EM) fields. Starting from the Lagrangian of the EM field:

$$\Lambda = -\frac{\varepsilon_0}{4} F^{\mu\nu} F_{\mu\nu} = -\frac{\varepsilon_0}{4} \left( g^{\alpha\mu} g^{\beta\nu} \right) F_{\alpha\beta} F_{\mu\nu} \qquad (1)$$

it was assumed that the space-time metric was a function of the 4-vector potential $\mathcal{A}$ from which the **EM** field derives, which makes $\mathcal{A}$ a real physical quantity[2].

Under these conditions, the coefficients $g^{\alpha\beta}$ are functions of the geometric coordinates $x^\nu$, on the one hand *via* the generalized coordinates $\mathcal{A}_\mu$, on the other hand due to possible other forms of energy (i.e. other sources of gravitation) or if the 4-space is planar and related to curvilinear coordinates. From the perspective of general relativity, the generality of Einstein's equations is not restricted. The gravitational field, that is to say the metric, is defined by all the forms of energy present. Simply, the part due to the EM field is, according to our assumption, defined by the field potentials.

We have deduced that Maxwell's equations of the 2nd group have, in this case, a generally non-zero second member and are written:

$$\varepsilon_0 F^{\mu\nu}_{;\nu} = \frac{1}{2} \tau_{\alpha\beta} \frac{\partial g^{\alpha\beta}}{\partial \mathcal{A}_\mu} \qquad (2)$$

where $\tau_{\alpha\beta}$ is the stress-energy tensor of the classical EM field and the $\partial g^{\alpha\beta}/\partial \mathcal{A}_\mu$, the derivatives of the contravariant components of the tensor metric with respect to the covariant components of the 4-potential vector $\mathcal{A}$ *on which they are assumed to depend*[3].

**1.2 The "Polarized Vacuum"**

Our hypothesis suggests an analogy with polarized matter. In a vacuum and in the presence of a gravitational field, the electrical **D** and magnetic **B** inductions are expressed as a function of the electrical **E** and magnetic **H** fields by the expressions ([2], §90):

$$\boldsymbol{D} = \frac{\varepsilon_0}{\sqrt{g_{00}}} \boldsymbol{E} + \frac{1}{c} \boldsymbol{H} \wedge \boldsymbol{g} \qquad \boldsymbol{B} = \frac{\mu_0}{\sqrt{g_{00}}} \boldsymbol{H} - \frac{1}{c} \boldsymbol{E} \wedge \boldsymbol{g} \qquad (3)$$

where **g** is the 3-vector of covariant components: $g_i = -g_{0i}/g_{00}$

---

[2] According to the classical theory of electromagnetism, except to introduce "gauge conditions" (like the Coulomb gauge), the 4-potential vector A is not uniquely defined. Its components are therefore not considered as physical quantities in the sense of measurable quantities. However, the vector potential is observable in quantum mechanics (Aharonov-Bohm effect).
[3] These derivatives are thus the components of a tensor of the 3rd order.

For weak EM fields, these expressions are linear since the gravitational field, if any, is given. In the absence of gravity (Minkowski space in an inertial reference frame[4]), they simplify to give the classical relations:

$$\boldsymbol{D} = \varepsilon_0 \boldsymbol{E} \quad \boldsymbol{B} = \mu_0 \boldsymbol{H} \qquad (4)$$

But if, as has been assumed for the extremely intense EM field, the gravitational field is affected by the 4-vector potential $\mathcal{A}$ from which the $\boldsymbol{E}$ and $\boldsymbol{H}$ fields derive, the relations (3) are no longer linear.

This situation can be compared to the case of polarized matter, where the EM field does not always obey the superposition principle. In a perfect dielectric, the electric induction $\boldsymbol{D}$ depends linearly on the electric field $\boldsymbol{E}$ but nonlinear phenomena occur if the field $\boldsymbol{E}$ is too high (disruptive discharge in gases, for example). The same is true for a paramagnetic substance cooled between the magnetic induction $\boldsymbol{B}$ and the magnetic field $\boldsymbol{H}$ (saturation phenomenon).

Drawing inspiration from this parallel, let us pose, for a material that is both dielectric and magnetic:

$$\boldsymbol{D} = \varepsilon_0 \boldsymbol{E} + \boldsymbol{P} \quad ; \quad \boldsymbol{B} = \mu_0 (\boldsymbol{H} + \boldsymbol{M}) \qquad (5)$$

where $\boldsymbol{P}$ and $\boldsymbol{M}$ would be, in a vacuum, the respective equivalents of the *polarization and magnetization intensity vectors* in matter, linked here to the $\boldsymbol{E}$ and $\boldsymbol{B}$ fields *via* the metric by the relations (3).

One can object that this formulation is devoid of physical meaning since, unlike a dielectric, the "polarization" does not involve any charge carrier such as an atom or a molecule. Under these conditions, monopoles are only fictitious quantities like magnetic masses, considered as a convenient representation of magnetized matter.

From our point of view, the monopoles created are, on the contrary, very real: these are *the usual electric charges and these are the manifestation of an extremely intense EM field*. Thus it is no longer necessary to consider the charge and the field as two independent physical entities and to study only their interactions, as classical theory does. Here, only the field exists and the term "charge" refers to any particular configuration of the field where the principle of superposition does not apply.

In classical theory, the charge is at the origin of the field but the reciprocal is not true. This is why, in our previous article [1], we limited ourselves to the case of the free field. Our hypothesis, in which the extremely intense EM field "creates" the electric charge, leads us now to consider the system "EM field + charge" as a single entity.

---

[4] And more generally, in a synchronous repository.

The right-hand sides of the equations represent the components of a 4-force f in the configuration space of the EM field, called, by analogy with the 4-density of electric current, *"4-density of electrogravitational current"* (or density of ecg) or:

$$-\frac{1}{c} f^\mu = \frac{1}{2} \tau_{\alpha\beta} \frac{\partial g^{\alpha\beta}}{\partial \mathcal{A}_\mu} \qquad (6)$$

Noting the formal identity of these equations with those, classical, of the field in the presence of charges, it was suggested, in the conclusion of our article, that the four-vector f could be identified with the 4-density of electric current *j*, this which is equivalent to saying: *"the 4-density of electric current j is the manifestation of the gravitational field created by an extremely intense EM field"*. Taking into account, its contravariant components are then written:

$$j^\mu = -\frac{c}{2} \tau_{\alpha\beta} \frac{\partial g^{\alpha\beta}}{\partial \mathcal{A}_\mu} \qquad (7)$$

### 1.3 Stress-Energy Tensor of the Intense EM Field

Under these conditions, it seems natural[5] to establish the stress-energy tensor of the intense EM field by deriving the Lagrangian (1) with respect to the geometric coordinates *via the covariant components of the 4-potential vector* $\mathcal{A}$. Here the coefficients depend on the geometric coordinates, on the one hand via the $\mathcal{A}_\mu$ (gravitational field due to the 4-potential), on the other hand directly (external gravitational field and/or curvilinear coordinates in Euclidean space):

$$\frac{\partial g^{\alpha\beta}}{\partial x^\nu} = A_{\mu,\nu} \left( \frac{\partial g^{\alpha\beta}}{\partial \mathcal{A}_\mu} \right)_{M\ const} + \left( \frac{\partial g^{\alpha\beta}}{\partial x^\nu} \right)_{\mathcal{A}\ const} \qquad (8)$$

which gives:

$$\tau_{\mu;\nu}^\nu = -\frac{1}{c} j^\nu F_{\mu\nu} \qquad (9)$$

The stress-energy tensor is then the sum of the classical EM field tensor:

$$\tau^{\mu\nu} = -\varepsilon_0 F^\mu{}_\alpha F^{\nu\alpha} + \frac{\varepsilon_0}{4} g^{\mu\nu} F_{\alpha\beta} F^{\alpha\beta} \qquad (10)$$

and a tensor π of components:

---

[5] This method is considered purely formal, its advantage being to directly obtain the stress-energy tensor in symmetric form.

$$\pi^{\mu\nu} = -\left(\mathcal{A}^\mu j^\nu + \mathcal{A}^\nu j^\mu - g^{\mu\nu}\mathcal{A}_\alpha j^\alpha\right) \quad (11)$$

or:

$$T^{\mu\nu} = \tau^{\mu\nu} + \pi^{\mu\nu} \quad (12)$$

The 4-divergence of its trace is identically zero provided that the quadrivector:

$$\frac{1}{c}\mathcal{A}^\nu\left(\partial_\mu j_\nu - \partial_\nu j_\mu\right) = 0 \quad (13)$$

which is the case in what follows.

## 2 Electrostatic Potential with Central Symmetry

In the reference article [1], we sought to determine the gravitational potentials generated by relatively weak EM fields. We addressed two simple cases: the constant electric field and the plane traveling wave. We are now interested in the case, *a priori* also simple, of the centrally symmetric electrostatic potential. The model must be a solution of the modified Maxwell equations (2) on the one hand, and of Einstein on the other.

The 4-vector potential $\mathcal{A}$ and the tensor $F$ of the EM field which derives from it are summarized for the first in the electrostatic potential $V = \mathcal{A}_0$ [6], which depends only on the distance to the center *r* and for the second on the radial electrostatic field $E = F_{01} = -\dfrac{dV}{dr}$

The gravitational field generated by the potential being centrally symmetric, we proceed as for the Schwarzschild and Reissner-Nordström metrics by setting:

$$g_{00} = e^\nu \; ; \; g_{11} = -e^\lambda \; ; \; g_{22} = -r^2 \; ; \; g_{33} = -r^2 \sin^2\theta \quad (14)$$

### 2.1 Maxwell's Equations

Since the system is not time dependent, the only non-zero Maxwell equation of the 2nd group (2) is:

$$\varepsilon_0 F^{01}_{;1} = -\frac{1}{c} j^0 \quad (15)$$

with:

$$F^{01}_{;1} = \frac{1}{\sqrt{-g}}\frac{d}{dr}\left(\sqrt{-g}\, g^{00} g^{11} F_{01}\right) = -\frac{1}{r^2 e^{\frac{\lambda+\nu}{2}}}\frac{d}{dr}\left(r^2 E\, e^{-\frac{\lambda+\nu}{2}}\right) \quad (16)$$

---

[6] The three-dimensional potentials here are the covariant components of the 4-potential: $\mathcal{A}_\mu = \left(\mathcal{A}_0 = V, \mathcal{A}_i = -c A_i\right)$

According to (7) and taking into account that only the components $g^{00}$ and $g^{11}$ depend on $V$ (the components $g^{22}$ and $g^{33}$ are Euclidean), $j^0$ thus has the expression:

$$j^0 = -\frac{c}{2}\tau_{\alpha\beta}\frac{\partial g^{\alpha\beta}}{\partial \mathcal{A}_0} = -\frac{c}{2}\tau_{\alpha\beta}\frac{dg^{\alpha\beta}}{dV} = -\frac{c}{2}\left(\tau_{00}\frac{dg^{00}}{dV} + \tau_{11}\frac{dg^{11}}{dV}\right) \quad (17)$$

From (10) and (14), we get:

$$\tau_{00} = \frac{\varepsilon_0}{2}E^2 e^{-\lambda} : \tau_{11} = -\frac{\varepsilon_0}{2}E^2 e^{-\lambda} \quad (18)$$

A priori, the derivatives $\frac{dg^{00}}{dV}$ and $\frac{dg^{11}}{dV}$ are not known but here we can write:

$$\frac{dg^{\alpha\alpha}}{dV} = \frac{dg^{\alpha\alpha}}{dr}\frac{dr}{dV} = -\frac{1}{E}\frac{dg^{\alpha\alpha}}{dr} ; \alpha = 1.2 \quad (19)$$

Which resolves to:

$$j^0 = \frac{c\varepsilon_0}{4}E\frac{de^{-\lambda-v}}{dr} = -\frac{c\varepsilon_0}{4}E(\lambda'+v')e^{-\lambda-v} \quad (20)$$

Hence the differential equation:

$$\frac{1}{r^2}\frac{d}{dr}\left(r^2 E e^{-\frac{\lambda+v}{2}}\right) = -\frac{1}{4}E(\lambda'+v')e^{-\frac{\lambda+v}{2}} \quad (21)$$

which integrates as follows:

$$\frac{\frac{d}{dr}\left(r^2 E e^{-\frac{\lambda+v}{2}}\right)}{r^2 E e^{-\frac{\lambda+v}{2}}} = -\frac{\lambda'+v'}{4} \Rightarrow r^2 E e^{-\frac{\lambda+v}{2}} = \text{Cste}*e^{-\frac{\lambda+v}{4}} \Rightarrow E = \frac{\text{Cste}}{r^2}e^{\frac{\lambda+v}{4}} \quad (22)$$

At a large distance from the origin the metric is Euclidean i.e. $\lambda + v \to 0$ and we must remember Coulomb's law, i.e. write the constant in the form $Q_t/4\pi\varepsilon_0$ where here $Q_t$ represents, according to our hypothesis, *the total charge generated by the field*. Hence the expression for the field:

$$E(r) = \frac{1}{4\pi\varepsilon_0}\frac{Q_t}{r^2}e^{\frac{\lambda+v}{4}} \quad (23)$$

Thus, due to the gravitational field it creates and *except in the case where $\lambda+v=0$*, the electric field does not obey Coulomb's law.

## 2.2 The Question of Containment

At the distance $r$ from the center, the charge $Q(r)$ is obtained by integrating the spatial charge density $\rho = \frac{1}{c}\sqrt{g_{00}}\,j^0$ in the volume of a sphere of radius $r$.

The elementary electric charge has the expression:

$$dQ(r) = \rho dv = \left(\frac{1}{c}\sqrt{g_{00}}\,j^0\right)\left(\sqrt{-g_{11}}\,4\pi r^2 dr\right) = \frac{4\pi}{c}j^0 e^{\frac{\lambda+v}{2}} r^2 dr \quad (24)$$

Considering (20) and (23), it follows:

$$dQ(r) = \frac{4\pi}{c}j^0 e^{\frac{\lambda+v}{2}} r^2 dr = \frac{4\pi}{c}\left[\frac{c\varepsilon_0}{4}\left(\frac{1}{4\pi\varepsilon_0}\frac{Q_t}{r^2}e^{\frac{\lambda+v}{4}}\right)\frac{de^{-\lambda-v}}{dr}\right]e^{\frac{\lambda+v}{2}} r^2 dr = Q_t\, de^{-\frac{\lambda+v}{4}} \quad (25)$$

From which:

$$Q(r) = Q_t e^{-\frac{\lambda+v}{4}} \quad (26)$$

*If the sum $\lambda+v$ depends on $r$, the electric charge is not confined.* Indeed, if as we have admitted, there is no boundary between the classical EM field (i.e. "weak" and for which the principle of superposition is verified) and the extremely intense EM field, then the associated electric charge must extend to infinity, because, however weak it is, the EM field creates gravitation.

Without undermining our approach, we will show that this is not the case and that the electric charge "created" by the field is necessarily confined in a sphere centered at the origin.

## 2.3 Einstein's Equations

In the present case, the only source of energy being electromagnetic; the stress-energy tensor is that of the EM field. Condition (13) holds since $\mu = v = 0$.

The components $\tau$ and $\pi$ of the stress-energy tensor result from expressions (10), (11), (12), (18) and (20), i.e.:

$$\tau_0^0 = \tau_1^1 = -\tau_2^2 = -\tau_3^3 = \frac{\varepsilon_0}{2}E^2 e^{-\lambda-v} \;;\; \mu \neq v \Rightarrow \tau_\mu^v = 0$$

$$\pi_0^0 = -\pi_1^1 = -\pi_2^2 = -\pi_3^3 = -\frac{1}{c}A_0 j^0 = \frac{\varepsilon_0}{4}EV(\lambda'+v')e^{-\lambda-v} \;;\; \mu \neq v \Rightarrow \pi_\mu^v = 0$$

(27)

Since the system does not depend on time, Einstein's equations are written ([3] §97):

$$-e^{-\lambda}\left(\frac{1}{r^2}-\frac{\lambda'}{r}\right)+\frac{1}{r^2}=\frac{8\pi G}{c^4}T_0^0$$

$$-e^{-\lambda}\left(\frac{v'}{r}+\frac{1}{r^2}\right)+\frac{1}{r^2}=\frac{8\pi G}{c^4}T_1^1 \qquad (28)$$

$$-\frac{1}{2}e^{-\lambda}\left(v''+\frac{v'^2}{2}+\frac{v'-\lambda'}{r}-\frac{v'\lambda'}{2}\right)=\frac{8\pi G}{c^4}T_2^2=\frac{8\pi G}{c^4}T_3^3$$

Subtracting the second from the first of these equations, it follows:

$$e^{-\lambda}\frac{\lambda'+v'}{r}=\frac{8\pi G}{c^4}(T_0^0-T_1^1)=\frac{8\pi G}{c^4}\left[\frac{\varepsilon_0}{2}EV(\lambda'+v')e^{-\lambda-v}\right]$$

$$=\frac{1}{V_p^2}EV(\lambda'+v')e^{-\lambda-v} \qquad (29)$$

where $V_p=c^2/4\pi\varepsilon_0 G$ is the Planck potential.

It follows that if $\lambda' + v' \neq 0$:

$$\frac{1}{r}=\frac{1}{V_p^2}EVe^{-v} \qquad (30)$$

from which:

$$g_{00}=\frac{1}{V_p^2}EVr \qquad (31)$$

But $g_{00}$ must trend to 1 when $r \to \infty$ at the same time the EM field must be Coulombic. It follows that $\lambda' + v' = 0$. Since the system is static and the metric is Euclidean at infinity, we deduce that:

$$\lambda+v=0 \qquad (32)$$

*Thus, the electric field (23) is reduced to the Coulombic field. We conclude that the limits of validity of the principle of superposition, if they exist, are thus hidden.*

### 2.4 The Reissner-Nordström Metric

The last two equations (28) form only one which become, for the first:

$$-\frac{1}{2}e^v\left(v''+\frac{2v'}{r}+v'^2\right)=-\frac{1}{2r^2}\frac{d}{dr}\left(r^2\frac{de^v}{dr}\right) \qquad (33)$$

and for the second:

$$\frac{8\pi G}{c^4}T_2^2 = \frac{8\pi G}{c^4}T_3^3 = \frac{8\pi G}{c^4}\left(-\frac{\varepsilon_0}{2}E^2\right) = -\frac{E^2}{V_p^2} = -\frac{1}{V_p^2}\left(\frac{1}{4\pi\varepsilon_0}\frac{Q_t}{r^2}\right)^2 \quad (34)$$

which leads to the differential equation:

$$\frac{d}{dr}\left(r^2\frac{de^\nu}{dr}\right) = \frac{2}{V_p^2}\left(\frac{1}{4\pi\varepsilon_0}\frac{Q_t}{r}\right)^2 \quad (35)$$

which integrates to provide:

$$e^\nu = \frac{1}{V_p^2}\left(\frac{Q_t}{4\pi\varepsilon_0}\right)^2\frac{1}{r^2} - \frac{r_0}{r} + \text{cste} = \frac{V^2}{V_p^2} - \frac{r_0}{r} + \text{cste} \quad (36)$$

The first integration introduces a radius $r_0$ to be determined. Moreover, when $r\to\infty$, $V\to 0$ and the metric is Euclidean $\Rightarrow g_{00} = e^\nu = 1 \Rightarrow \text{cste} = 1$. Finally, since the potential is Coulombic: $\frac{V}{V_p} = \frac{r_p}{r}$ from which:

$$g_{00} = -g^{11} = e^\nu = 1 - \frac{r_0}{r} + \frac{r_p^2}{r^2} \quad (37)$$

These are the components of the Reissner-Nordström metric ([5], p.840-841)

$$ds^2 = \left(1 - \frac{r_S}{r} + \frac{r_Q^2}{r^2}\right)c^2 dt^2 - \frac{1}{1 - \frac{r_S}{r} + \frac{r_Q^2}{r^2}}dr^2 - r^2 d\Omega^2 \quad (38)$$

which usually describes the gravitational field created in its proximity by a massive electrically-charged and immobile body (a star, a black hole):

$r_S$ is the Schwarzschild radius, proportional to its mass M:

$$r_S = \frac{2GM}{c^2} \quad (39)$$

$r_Q$ is a length proportional to its charge Q:

$$r_Q^2 = \frac{G}{4\pi\varepsilon_0 c^4}Q^2 \quad (40)$$

The condition $g_{00} \geq 0$ assumes that $r_Q < r_S/2$ except when allowing for the possibility of bare singularities[7]. In reality, $r_Q \ll r_S$ which brings us back to stars lacking charge. The present case, however, differs from that of a star in that there is no mass here (in the classic sense of the term) determining a horizon. Moreover, except to decree that $r_0 = 0$ nothing here prohibits that $rp > r_0 / 2$. It's notably the case of the electron for which $r_S \sim 10^{-57}$m $\ll r_Q \sim 10^{-36}$m. Therefore, *this hypothesis is adopted for the models proposed later for the electron and dark energy.*

In this case $g_{00} > 0 \forall r$ and passes through a minimum for:

$$r_{0g} = 2\frac{r_p^2}{r_0} \tag{41}$$

It follows that by replacing in (37) $r_0$ with its expression taken from (41), the components of the tensor metric now have the expressions:

$$g_{00} = -g^{11} = 1 - \frac{r_p^2}{r}\left(\frac{2}{r_{0g}} - \frac{1}{r}\right) \tag{42}$$

with:

$$r_p < r_{0g} \tag{43}$$

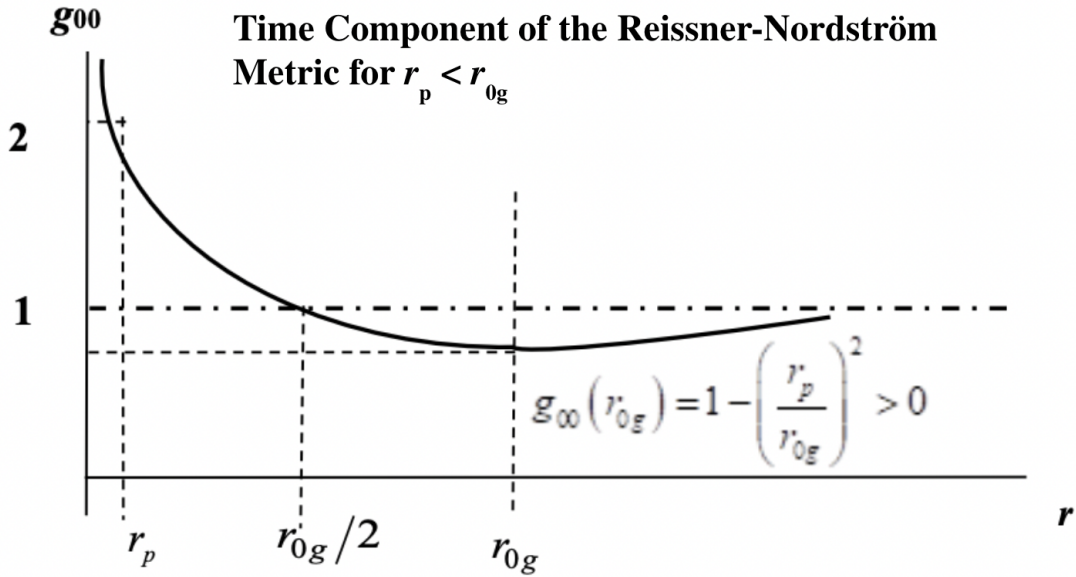

**Time Component of the Reissner-Nordström Metric for $r_p < r_{0g}$**

$$g_{00}(r_{0g}) = 1 - \left(\frac{r_p}{r_{0g}}\right)^2 > 0$$

---

[7] There is no consensus among astrophysicists about their existence due to Roger Penrose's "cosmic censorship principle" which is generally believed to be true. British physicist and mathematician, Roger Penrose is the author of a conjecture according to which there is no physical process giving rise to a bare singularity, that is to say a region of space whose gravitational field takes on infinite values and which would not be "hidden" behind an "event horizon".

## 2.5 Gravitational Acceleration

At zero speed, the only non-zero component of the 4-acceleration is radial and has the expression:

$$\frac{d^2r}{ds^2} = \frac{d^2x^1}{ds^2} = -\Gamma^1_{00}\left(\frac{dx^0}{ds}\right)^2 = -c^2\frac{v'}{2}e^{2v}\left(\frac{dt}{ds}\right)^2 \qquad (44)$$

Given that $\dot{r} = 0$:

$$\frac{d^2r}{ds^2} = \ddot{r}\left(\frac{dt}{ds}\right)^2 \qquad (45)$$

Hence the gravitational acceleration:

$$\ddot{r} = -c^2\frac{v'}{2}e^{2v} = -\frac{c^2}{2}e^v\frac{de^v}{dr} = -\frac{c^2}{2}g_{00}\frac{dg_{00}}{dr} \qquad (46)$$

which cancels out at the minimum of $g_{00}$, justifying the notation of $r_{0g}$ in (41).

Given that $g_{00} > 0 \forall r$, the gravitation is repulsive $(\ddot{r} > 0)$ for $r < r_{0g}$ and attractive $(\ddot{r} < 0)$ beyond. Thus, *a "bump" of the electric field is characterized by the electric charge created and by the distance to the center, where the induced gravitation is canceled while passing from repulsive to attractive.*

The gravitational potential is deduced by integrating (46). Given that it must cancel out to infinity, it follows:

$$\varphi(r) = \frac{c^2}{4}\left(g_{00}^2 - 1\right) \qquad (47).$$

## 2.6 Equivalent Mass

At a large distance the gravitational acceleration (46) has the following expression:

$$\ddot{r} \approx -c^2\frac{r_p^2}{r_{0g}}\frac{1}{r^2} \qquad (48)$$

a relationship that must be compared to Newton's law $\ddot{r} = -\frac{Gm_0}{r^2}$ to define the equivalent of the resting mass M in the metric of Reisser-Nordström (here denoted $m_0$).

$$m_0 c^2 = Q_t V_{0g} \qquad (49)$$

produces charge by electrostatic potential where gravitation cancels out.

## 3 EM Potential Expressed in Planck Units

### 3.1 A Limited Potential

Without a horizon like that of Schwarzschild since $g_{00} > 0 \forall r$, the electrostatic potential should always be Coulombic and the charge point concentrated at the origin (bare singularity). In fact, this circumstance makes it possible to deduce that *the charge, considered as a bump in the electric field, is necessarily confined* in accordance with our hypothesis of the gravitational nature of electric charges.

From our point of view, it would be *a priori* surprising if the principle of superposition remains valid whatever the amplitude of the field. Such a situation would contradict our hypothesis that the EM field no longer obeys the principle of superposition for very high EM potentials. This is why *we are led to suppose that the potential is Coulombic only beyond a certain distance $r_p$ from the center and that it is no longer below. In this central zone, the condition $\lambda + \nu = 0$ is therefore no longer respected*, which, given relation (20), means that it is charged. Thus *the charge, considered as a bump in the electric field, is necessarily confined, without being point-like*. Therefore, apart from the volume occupied by the charge, the classical theory (conservation of charge and gauge invariance) remains valid.

*Thus the charge would be contained in a central sphere whose radius would be determined by an unsurpassable electric potential which, as such, appears as a new physical constant.*

*In the general case, this limit potential can only be an invariant if it is a scalar, in this case the norm of the 4-potential $\mathcal{A}$, and not only its temporal component V. This norm then appears as a limiting physical quantity such as the speed of light.*

**Note:** Although this is not a device for avoiding a bare singularity, the introduction of a limit potential makes it possible to satisfy Roger Penrose's principle of cosmic censorship, already mentioned. In our case, there is no horizon(s) as for a Reissner-Nordström black hole but an equipotential delimits an EM bump.

### 3.2 What Potential?

In the absence of a determining criterion, we propose the Planck potential $V_p$: its value of approximately $10^{27}$ volts is compatible with our hypothesis that gravitation of EM-origin only appears for extremely high potentials, in the equations of General Relativity (GR) and for $r \geq r_p$, it naturally appears as the stress-energy tensor factor of the EM field:

$$\left| R_{(\mu)}^{(\mu)} \right| = \frac{8\pi G}{c^4} \left( \frac{1}{2} \varepsilon_0 E^2 \right) = \left( \frac{E}{V_p} \right)^2 \tag{50}$$

### 3.3 Inside the Bump

### 3.3.1 No Static Solution

Given that the quantity $\lambda' + \nu'$ is non-zero in the central zone, the component $g_{00}$ is given by an expression which can be written as:

$$g_{00} = -\frac{1}{2V_p^2} r \frac{dV^2}{dr} \qquad (51)$$

which shows that $g_{00} > 0$ provided that $|V|$ decreases as a function of the distance from the center. But, unless we allow a discontinuity in potential $V$ for $r = r_p$, this is contrary to the hypothesis that the potential $V$ is maximum in absolute value for $r = r_p$.

To define a metric in the central zone, we can use the fact that the components $g_{00}$ and $g_{11}$ are linked to the charge by the equivalent expression:

$$g_{00} g_{11} = -\frac{1}{q^4}; \quad q(r) = \frac{Q(r)}{Q_t} \qquad (52)$$

From which:

$$g_{11} = -e^\lambda = -\frac{1}{q^4} e^{-\nu} \qquad (53)$$

Which leads to the metric:

$$ds^2 = e^\nu c^2 dt^2 - \frac{1}{q^4} e^{-\nu} dr^2 - r^2 d\Omega^2 \qquad (54)$$

where $\nu$ and $q$ are functions of $r$ and $t$, via the 4-potential $\mathcal{A}$. This expression, which generalizes the one valid above the horizon for which $q = 1$, explains the representation by the electric charge of the gravitational field created by the EM field in the central zone[8].

On this basis Einstein's equations, combined with the field expression (23), should make it possible to determine $q$ $r$ and therefore the potential $V$ and the coefficients of the metric. But there are fewer variables than equations. None of them are the result of the others, there is no solution that satisfies them all. It follows that: *"below the horizon, the potential does not admit a static solution in the Reissner-Nordström metric"*.

It is possible that solutions exist, notably stationary in a soliton-type approach. It is also possible that it is futile to look for a "classical" model, as we will see later in the case of the electron.

---

[8] We note that we have: $\sqrt{-g} = q^2 \sqrt{-g^0}$ where $g^0$ is the determinant of the Reissner-Nordström metric.

### 3.3.2 Internal Energy

The rest mass $m_0$ or "equivalent mass" (49) is the sum of the electrostatic energy of the external field (Coulomb) and the energy contained in the sphere of radius $r_p$ which can be called "internal energy".
Hence the relation $V_p$ designating the potential on the surface of the sphere:

$$m_0 c^2 = \mathcal{E}\ r_p + \frac{1}{2} Q_t V_p \tag{55}$$

Replacing $m_0$ with its expression (49), we get:

$$\mathcal{E}\ r_p = -Q_t \left( \frac{1}{2} V_p - V_{0g} \right) = -\frac{1}{2} \frac{Q_t^2}{4\pi\varepsilon_0} \left( \frac{1}{r_p} - \frac{2}{r_{0g}} \right) \tag{56}$$

Given that $r_p < r_{0g}$, $\mathcal{E}\ r_p < 0$
In the case of a star (a spherical body of radius $r$), charged or not, its energy is given by its stress-energy tensor ([3], §97):

$$mc^2 = \mathcal{E}(r) = 4\pi \int_0^r T_0^0 u^2 du \tag{57}$$

where $T_0^0$ is $>0$. But there is no difficulty in assuming that the internal energy is $<0$ if it is purely EM in nature. Indeed, this mass is defined by the stress-energy tensor of the intense EM field. If the component $\tau_0^0$ is always $>0$, the same cannot be said for $\pi_0^0$ below the surface, so the internal energy may well be $<0$. Indeed, in configuration space, $\tau_0^0$ represents the "kinetic" energy density and $\pi_0^0$ the "potential" energy density of the EM field. To say that the total energy is $<0$ is to say that the potential energy is $<0$ and thus prevails, in absolute value, over the kinetic energy.

## 4 An Elementary Charge Model

### 4.1 Why the electron?

On the macroscopic scale, all material bodies have a mass independent of any charge they might have. It follows that a body, whose mass would be purely EM, can only be of subatomic dimensions, that is to say a particle.
The question is which particles can be candidates? The only charged particles, stable in the free state, are the electron[9] and the proton (and their antiparticles[10]). But the proton is a composite particle of known size and composition while the electron (or the positron) is deemed "elementary", that is to say without internal structure and even without dimensions[11].

---
[9] Lifetime $> 4.6 * 10^{26}$ years.
[10] From the electrogravitational point of view, they are indistinguishable because only the absolute value of the charge counts.

This is why we propose to describe it as a "bump" of pure EM energy, *a micro-sphere with a very small but non-zero radius*.

However, let us emphasize the limits of the exercise. At this scale only quantum mechanics apply. Furthermore, there is no evidence that the classical laws of gravitation apply. On the one hand, currently they are only experimentally validated above 50 μm, on the other hand, we are seeking to develop a quantum theory of gravitation as for the three other fundamental interactions[12] or better a global theory unifying them all.

But nothing requires that gravitation must be quantified at this scale. Furthermore, quantum effects require interaction with a measuring device. In the present case the charged particle is isolated (necessary for the Reissner-Nordström metric to apply) and we are only interested in some of the parameters which define it (mass, charge, dimension). This is why, while remaining cautious, the representation of a charged particle by a "classical" gravitational field does not seem unreasonable, nor does representing it by a charged point mass, which has no physical meaning.

In what follows we present some characteristics and consequences of our electron model.

**4.2 Radius of the "classical" electron**

In this case the Planck potential corresponds to the radius $r_e$ of the sphere which holds the elementary charge $e$, i.e.:

$$r_e = \frac{1}{4\pi\varepsilon_0}\frac{e}{V_p} \sim 1.4*10^{-36}\text{m} \tag{58}$$

which we believe represents the "classical electron's radius", not to be confused with the "classical radius of an electron" which corresponds here to the radius 0-g according to (49):

$$r_{cl} = \frac{e^2}{4\pi\varepsilon_0}\frac{1}{m_e c^2} \tag{59}$$

The radius $r_e$ of the electron (59) is related to Planck length $l_p$ by the relation:

$$r_e = l_p\sqrt{\alpha} \sim 0.085 l_p \tag{60}$$

---

[11] In accordance with relativity where an elementary particle is a non-deformable material object, therefore a geometric point where all the mass is concentrated. Furthermore, it is experimentally proven that the radius of the electron is $< 10^{-22}$ m.

[12] There is, for the moment, no sufficiently validated theoretical or experimental means of treating gravitation at this scale. The most recent measurements do not go below 50 μm and the theories of strings and loop quantum gravity are still in development.

where $\alpha$ is the fine-structure constant. It is thus smaller than the Planck length, generally considered to be a limit to observability.

For $r = r_e$, the non-zero components $R_{(\mu)}^{(\mu)}$ of the Ricci tensor (50), equal in absolute value, have for expression the Gaussian curvature of the sphere bounding the electron:

$$\left|R_{(\mu)}^{(\mu)}\right| = \frac{1}{r_e^2} \tag{61}$$

However, we cannot deduce an upper limit to the scalar curvature of 4-space on the grounds that the Planck potential cannot be exceeded. According to our hypothesis, the latter is a relativistic invariant while the radius $r_e$ is linked to our electron model (static with central symmetry)[13].

### 4.3 A Gravitational Cliff

On the surface of the electron, the gravitational (repulsive) acceleration reaches the phenomenal value of $2.6*10^{53}$ m*s$^{-2}$! But it drops very quickly (potential at $1/r^2$) and becomes completely negligible beyond $10^{-18}$ m, well before the classic radius $r_{0g}$.

## 5 A Model for Dark Energy (DE)

### 5.1 Repulsive Gravity

The acceleration of the expansion of the Universe, highlighted in the 1990s, supposes an anti gravitational effect opposing the attraction between and within the other components of the universe (gas clouds, galaxies and the hypothetical "dark matter"), which tends on the contrary to slow down the expansion. It appears that after a period of slowdown, acceleration began about 6 Gyr ago[14].

The most popular current hypothesis, although quite strange, is to add into the standard cosmological model a "cosmological fluid", called "dark energy" or "black" (Dark Energy or DE), supposed to fill the entire universe[15] and representing a preponderance of its total energy. Its density, noted $\varrho_\Lambda$, is assumed to be uniform and constant over time (or varying very slowly).

---

[13] As part of work on quantum gravity, the Italian physicist Giovanni Amelino-Camelia proposed in 2002 a modified theory of special relativity (known as DSR for Doubly Special Relativity) where it is posited, among other things, that the Planck length is a relativistic invariant. This could be the case of the $r_e$ radius which is connected to it. However, DSR is generally considered speculative, not validated to date.

[14] The two phases approximately compensate each other. It follows that the current radius of the universe ($R_u \sim 1.31*10^{26}$ m) is approximately the Hubble distance ($d_H \sim 1.37*10^{26}$ m).

[15] The European Space Agency (ESA) plans to launch the Euclid space telescope in 2022 to determine the origin and source of the accelerated expansion.

Currently it represents about 68% of the average energy density of the universe[16], or $6*10^{-27}$ kg*m$^{-3}$.

*Its essential property is to have a pressure < 0, which gives it an anti-gravitational character.* It does not correspond to any currently known particle and does not seem to interact with anything. This is why it is called "vacuum energy". Nevertheless, it seems difficult to equate it with the energy of the quantum vacuum from which it differs by more than 100 orders of magnitude.

## 5.2 The ΛCDM model

According to this model, the vacuum cannot expand (sic), it is accepted that the density of the DE is constant over time and therefore does not depend on the scale factor $\alpha$. Given that $\varrho_\Lambda \propto \alpha^{3(1+w)}$ [17], it follows that $w = -1 \Rightarrow \varrho_\Lambda = -\varrho_\Lambda$. The simplest stress-energy tensor which meets this characteristic consists of reintroducing[18] the "cosmological constant" which originally appeared in the 1st of Einstein's equations and passing it to the 2nd leads to the expression:

$$T_{\mu\nu}^{(\Lambda)} = g_{\mu\nu} \frac{c^4}{8\pi G} \Lambda \qquad (62)$$

For this reason, this cosmological model including DE is called "ΛCDM Universe" (Lambda Cold Dark Matter).

## 5.3 wCDM models

Measurements of the parameter $w$ give values around –1 [19], hence a density of the DE which can vary slowly over time. Note that below -1, it increases instead of decreasing. This exotic case is why DE is then called "phantom energy".
The cosmological constant would then only be an approximation[19]. As a result, many models have been proposed, grouped under the name "wCDM Universe". In particular those called quintessence[20] characterized by a scalar field.

---

[16] The average density of the universe is very close to the density, called critical, $\rho_c = \frac{3H_0^2}{8\pi G} \approx 8.6*10^{-27}$ kg*m$^{-3}$ where $H_0$ is the Hubble constant.

[17] $w$ designates the ratio between the pressure and the density of the DE.

[18] Introduced by Einstein to "freeze" the universe which was thought to be static, it was removed when expansion was discovered by Hubble.

[19] More generally, for there to be an acceleration of expansion, it is enough that (1st Friedmann equation):
$$\rho_T + 3p_T < 0 \Leftrightarrow (1+3w_T)\Omega_T < 0 \Rightarrow w_T < -\frac{1}{3}$$
Note that this is a strict inequality. There is no acceleration in case of equality.

[20] Quintessence: ethereal substance that certain ancient philosophers added as a fifth element to the traditional four elements.

In the ΛCDM model, "vacuum energy" gives meaning to the cosmological constant whose existence it justifies. From this point of view, quintessence may seem to be a better concept although the physical nature of the associated scalar field is also unknown. In fact, creating such a field *ex nihilo* on the sole basis of the known characteristics of the DE is ultimately hardly more convincing than introducing the cosmological constant.

Different paths are explored such as adapting/modifying current theories (GR, MOND[21]) or proposing new particles not yet detected (as for dark matter with the famous "WIMPs"[22]), etc. But some astrophysicists believe that DE cannot be understood within the framework of current physics. Thus, DE could be the opportunity for a decisive breakthrough with the objective of encompassing all physical phenomena in a single conceptual framework [23].

In this very speculative field, all ideas are interesting. Let's dare... go *off-piste*!

### 5.4 Electrostatic bumps

*Let us assume that the cosmic vacuum is a bumpy gas of the EM field.* Here we join the hypothesis that the DE would be none other than the energy of the quantum vacuum although their density is in a ratio $10^{126}$! However, recent work [21] could remove this difficulty by differentiating the average energy of the quantum vacuum from that of its fluctuations, which would be of the same order. We believe these fluctuations would be bumps in the EM field according to the model above.

By analogy, it is a 3D version of the 2D situation of a sailor in the open ocean. As long as no land is visible, regardless of where he is and the direction in which he looks, the sailor sees only an expanse, heterogeneous on the scale of waves, but uniform on the scale of the horizon.

Let's try to characterize these bumps. For simplicity, let's assume they are all identical (same charge |q|, same mass m) and, *on average*, equally distributed. A "trough" (-q) follows a "crest" (+q), so that *the medium is a homogeneous and electrically neutral plasma*.

The density of bumps is determined by the energy density of the DE at ∼5.4 × $10^{-10}$ J/m³ divided by the energy of a bump, i.e.:

$$n_b = \frac{\rho_{DE}}{mc^2} \qquad (63)$$

Dark energy is assumed to be cold, so the mass of a bump is almost its mass at rest given by (49), hence:

$$n_b \sim \frac{\rho_{DE}}{qV_{0g}} = \frac{4\pi\varepsilon_0}{q^2}\rho_{DE}r_{0g} \sim 6*10^{-20}\frac{r_{0g}}{q^2} \text{ bumps / m}^3 \qquad (64)$$

---

[21] For "**MO**dified **N**ewtonian **D**ynamics". Modification of Newton's theory proposed in 1983 by the Israeli physicist Mordehai Milgrom to explain the too rapid rotation of stars and galaxies.
[22] For "Weakly Interacting Massive Particles".
[23] GR and QM are the two basic tools of cosmology. They complement each other but remain incompatible despite the considerable efforts made to unify them.

Thus, *the density of bumps is proportional to their radius 0-g and inversely proportional to the square of their charge*.

In order for this gas of bumps to generate an ubiquitously repulsive gravitational potential, the average distance $\bar{d} = n_b^{-1/3}$ between bumps must be < radius 0-g of the bumps:

$$\bar{d} = \left(4\pi\varepsilon_0 \rho_{DE} \frac{r_{0g}}{q^2}\right)^{-1/3} < r_{0g} \Rightarrow r_{0g} > \left(\frac{q^2}{4\pi\varepsilon_0 \rho_{DE}}\right)^{1/4} \sim 6.4*10^4 \sqrt{q}\ \text{m} \quad (65)$$

Considering that the bumps are elementary particles, we can assume that their charge is as well and set $q = e$. Thus:

$$r_{0g} > \left(\frac{e^2}{4\pi\varepsilon_0 \rho_{DE}}\right)^{1/4} \sim 2.6*10^4\ \text{m} \quad (66)$$

that is, $\sim 10^{-10}$ the mass of the electron.

So, according to our model, dark energy would be neutral, cold, and rarefied plasma ($n_b \sim 10^{13}$ bumps/m³), made up of particles of charge |e| and very low mass (<$10^5$ eV), or even zero. Naturally, it remains to be proven that such particles exist [24].

### 5.5 A cosmological fluid

Given the supposed EM nature of DE, we must be able to establish the equation of state $p_{DE} = w\rho_{DE}$ which characterizes $(p_{DE} \approx -\rho_{DE} \Leftrightarrow w = -1)$ based on our EM theory. We assume the parameter $w$ is constant over time.

The 4-space is related to co-moving coordinates and provided with the Robertson-Walker metric with $k = 1$ (in flat space):

$$ds^2 = c^2 dt^2 - a^2\left(\frac{dr^2}{1-r^2} + r^2 d\Omega^2\right) \quad (67)$$

The equation of state is established from the stress-energy tensor $T^{\mu\nu} = \tau^{\mu\nu} + \pi^{\mu\nu}$ of the EM field, given by the relations (10) and (11). The energy density is given by its time component $\varrho_e = T^{00} = \tau^{00} + \pi^{00}$, or:

$$\tau^{00} = \frac{1}{2a^2}\left(\varepsilon_0 E^2 + \mu_0 B^2\right) \quad (68)$$

and:

---

[24] Known particles of very low or zero mass (neutrinos, gluons, photons) are not charged. The same is true of the hypothetical graviton.

$$\pi^{00} = -\frac{1}{c}\left[2\mathcal{A}^0 j^0 - g^{00}\left(\mathcal{A}_0 j^0 + \mathcal{A}_i j^i\right)\right]$$
$$= -\frac{1}{c}\left(\mathcal{A}^0 j^0 - \mathcal{A}_i j^i\right) = -(\rho V + jA) \tag{69}$$

From which:

$$\rho_e = \tau^{00} - (\rho V + jA) \tag{70}$$

The pressure is deduced from the stress-energy tensor trace:

$$T = \pi^\mu_\mu = -\frac{1}{c}\left(2\mathcal{A}_\mu j^\mu - \delta^\mu_\mu \mathcal{A}_\alpha j^\alpha\right) = \frac{2}{c}\mathcal{A}_\mu j^\mu$$
$$= \frac{2}{c}\left(\mathcal{A}_0 j^0 + \mathcal{A}_i j^i\right) = 2(\rho V - jA) \tag{71}$$

From which:

$$p = -\frac{1}{3}(T - \rho_e) = -\frac{1}{3}\left[2(\rho V - jA) - \tau^{00} + (\rho V + jA)\right]$$
$$= \frac{1}{3}\left[\tau^{00} - (3\rho V - jA)\right] \tag{72}$$

In summary:

$$\rho_e = \tau^{00} - (\rho V + jA)$$
$$p = \frac{1}{3}\left[\tau^{00} - (3\rho V - jA)\right] \tag{73}$$
$$T = 2(\rho V - jA)$$

We deduce the equation of state:

$$p = w\rho_e \Rightarrow \frac{1}{3}\left[\tau^{00} - (3\rho V - jA)\right] = w\left[\tau^{00} - (\rho V + jA)\right] \tag{74}$$

Or:

$$\left(w - \frac{1}{3}\right)\tau^{00} - (w-1)\rho V - \left(w + \frac{1}{3}\right)jA = 0 \tag{75}$$

equality which determines, knowing $w$, one of the three terms in play from the two others. (73) becomes, by eliminating $\tau^{00}$ and in the case where $w \neq \frac{1}{3}$:

$$\rho_e = -\frac{2}{3w-1}(\rho V - jA) \qquad (76)$$

The case where $w = \frac{1}{3}$ corresponds to classical EM radiation for which $\rho V = jA = 0$ and thus: $\varrho_\gamma = \tau^{00}$; $T = 0$.

### 5.6 The case of Dark Energy

For $w \approx -1$ it follows:

$$\rho_{DE} \approx \frac{1}{2}(\rho V - jA) \sim -p_{DE} \qquad (77)$$

In this form, the DE is determined solely by the quantity $\rho V = jA$ with the condition $\rho V > jA$, the energy density and the trace of the stress-energy tensor must be > 0.
The average of $\varrho$ is zero due to the neutrality of the bump plasma. However, the product $\varrho V$, proportional to the square of the charge of the bumps, is > 0. Furthermore, the velocities of the bumps being low, $j \sim 0$ everywhere so that $\varrho_\Lambda \sim \varrho V/2$, the classical expression of the electrostatic energy density of a system of charges.
*Thus, the proposed model seems capable of representing a cosmological fluid with negative pressure, characteristic of DE.*

## 6 Discussion

### 6.1 The Bumpy EM Field

First Hendrik Antoon Lorentz around 1890 then Henri Poincaré in 1905 imagined the electron as a small packet of purely EM energy. Gustav Mie in 1912, and later Max Born and Leopold Infeld in 1934, proposed the concept of a "field of bumps" according to which the particles would be "*extended singularities of a unified field U governed by a system of equations to nonlinear partial derivatives which are deduced from a variational principle associated with a Lagrangian L*". As for Einstein, he thought that the division between matter and field was artificial and that it was only a difference in the concentration of energy, that of matter being very much greater than that of the external field. Thus "*a thrown stone is a changing field in which the states of the greatest field intensity travel through space with the velocity of the stone*[25].

---

[25] The "wave packet" associated with a particle in quantum mechanics can be interpreted in this way although they are probability waves.

The laws governing EM fields lead, in our view, to the concept of a field of bumps. The electric charge is none other than the "field bump" created by the gravitational effect of a locally extremely intense EM field.

This conjecture has the advantage of providing a "classical" electron model that is more relevant than its reduction to a material point endowed with a mass and a charge.

### 6.2 Electromagnetism and gravitation

It is surprising that the concept of a "field of bumps" was not used in its time. We can think that the revolutionary concept of a space-time sculpted by matter aroused the hope of unifying gravitation and electromagnetism by generalizing the geometric approach of the GR to the detriment, perhaps, of a more physical conception of the field.

By identifying the 4-current density of Maxwell's equations in the presence of charges with the "4-electrogravitational current density", we achieve a form of unification of electromagnetism and gravitation, at least in a classical framework. This may seem pretentious, Einstein and his contemporaries having failed in this attempt. We can also say that it comes too late insofar as the current objective is to unify all known interactions. Finally, we can object that its practical interest is non-existent, since on a large scale classical theory is sufficient and on the Planck scale only quantum physics is operative.

On this scale nothing proves, in fact, that the classical laws of gravity apply. On the one hand, for the moment, they are only experimentally validated above 50 μm, on the other hand we have been seeking for decades to develop a quantum theory of particles and fields including gravitation, without real success for the moment. It is also not certain that gravitation must be quantified at this scale [22].

### 6.3 Spin of the electron

Since the "classical" electron is not point-like, we can think of attributing to it its own angular momentum and therefore establish a link with its spin, a purely quantum quantity with no classical equivalent in the point representation. The associated metric would then be that of Kerr-Newman. It is possible that this approach makes it possible to find a solution to the problem of its internal structure as posed previously.

### 6.4 Dark Energy

The representation of DE by a gas of quasi-point bumps is probably too simplistic. Rather than points, the peaks of potential could form more or less sinuous lines, like the crests of waves.

However, this should not call into question the essential fact that the gravitational field created by the bumps can be repulsive everywhere[26].

The fact that the DE is undetectable (at least until now) other than by its gravitational effects could be explained by the fact that the bumps which constitute it interact only very weakly with each other or with the other components of the DE. universe. This "perfect gas" which would fill the entire universe would have the same properties as the "vacuum" of the cosmological constant while giving it its substance (quintessence).

**7 Conclusion and Perspectives**

In the absence of experimental evidence, the EM field bump concept and the resulting electron and DE models are eminently refutable (starting with basic theory). In any case, these models are certainly too simplistic and it is possible that they contain errors, perhaps fatal, of principle or calculation. Here EM bumps, supposed to be at the origin of DE, would be charged particles with almost zero mass, which does not correspond to any particle known or believed to exist[27].

It seems to us, however, that our approach has some interest:
- It rests on a known basis, in this case the EM field, of which we assume only that the principle of superposition no longer applies for extremely high potentials and fields, very far, it is true, from those that we observe or know how to produce.
- The classical EM field and the elementary charge are no longer independent entities, as is the case in classical electromagnetism.
- It is supported by the fact that any form of energy generates gravitation, provided it is appreciable, which, in our opinion, extremely intense EM potentials and fields should allow.
- It proposes a non-point "classical" electron model, a more realistic concept than the currently accepted dimensionless particle.
- It offers a model which seems adapted to DE without it being necessary to postulate the existence of a new field with *ad hoc* properties.

That being said, *our interest lies in the antigravitational character of the space surrounding the bumps*. If proof could be provided, it would perhaps be the start of a major scientific and technological revolution, particularly in the field of space transport. This perspective is at the origin of our thesis.

---

[26] Without having demonstrated it in a general way, this property appears in other cases so that we can think that extremely high EM fields locally produce an anti gravitational field.

[27] Unlike the hypothetical WIMPs, supersymmetric, massive, and neutral particles, supposed to constitute dark matter, which is attractive.